\documentstyle[12pt]{article}
\title{Comment on ``Spontaneous Inflation and the Origin 
of the Arrow of Time"}
\author{Hrvoje Nikoli\'c \\
Theoretical Physics Division, Rudjer Bo\v{s}kovi\'{c} Institute, \\
P.O.B. 180, HR-10002 Zagreb, Croatia  \\
{\normalsize e-mail: hrvoje@thphys.irb.hr} \\
}
\begin{document}
\maketitle
\vspace*{0.5cm}
\begin{abstract}
Recently, Carroll and Chen [hep-th/0410270] suggested a 
promising natural explanation of the thermodynamic arrow of time. 
However, we criticize their assertion that there exists 
a Cauchy hypersurface with a minimal entropy and argue
that such a Cauchy hypersurface is not needed for an explanation 
of the arrow of time.
\end{abstract}
\vspace{1cm}

\noindent
Recently, Carroll and Chen \cite{car} proposed an 
interesting natural 
explanation of the thermodynamic arrow of time. 
Their proposal consists of two main ingredients. 
First, 
they observe that if an arbitrarily small patch of space 
contains an {\em infinite} number of the degrees of freedom, 
then the entropy of this patch is not bounded from above, 
so it is not unnatural to have a state in which the entropy is 
not the maximal possible one. This allows the entropy to 
always increase from {\em any} given starting point. 
Second,
in order to understand why the process of entropy creation 
would create regions of spacetime resembling our observable universe, 
they appeal to spontaneous eternal inflation requiring 
a positive cosmological constant and an appropriate inflaton field.

In this comment, we have little to say regarding the second 
ingredient related to inflation. Instead, our comment
refers to the first ingredient, which is the essential one for 
the understanding of the time arrow itself. Since the 
two opposite directions of time should, {\it a priori}, 
play equal roles, Carroll and Chen argue that for a 
{\em typical} initial condition on a given Cauchy 
hypersurface, the entropy increases in {\em both} 
directions from this hypersurface. Thus, they predict the 
existence of a turning point (or, more precisely, of a 
turning Cauchy hypersurface) at which 
the entropy attains a minimum. (Incidentally, the existence 
of such a turning Cauchy hypersurface was suggested also in 
\cite{nik}, but there, such a hypersurface was considered 
special, rather than typical.) In this comment, we 
criticize the assertion of Carroll and Chen that, for a 
typical initial condition, the entropy increases in 
both directions from the initial hypersurface. Instead, 
we suggest a reformulation of their proposal, so that 
such a turning hypersurface is not needed for an explanation 
of the arrow of time.

Our criticism goes as follows. According to the picture  
of the universe proposed in \cite{car}, the entropy
increases in both time directions from the 
initial hypersurface. However, at {\em any} other Cauchy
hypersurface (that does not share common points with the initial one)
the entropy increases typically {\em only in one} direction.
But this means that, among all hypersurfaces, the initial
hypersurface has a {\em unique} property of having the time arrow
in both directions. In other words, the initial
hypersurface having two directions of time 
is {\em not typical at all}.
Instead, {\em the points of a typical hypersurface have 
a time arrow in one direction only}.

Another way of formulating our objection is as follows.
Let us consider the whole solutions of the equations of 
motion, rather than the initial conditions.
Given the assumption that the entropy is
unbounded from above, what are the properties of  
typical solutions defined 
{\em everywhere} in the universe? We agree with the 
assessment of \cite{car} that most, if not all, of the 
spacetime points in the universe will have a time arrow. 
But is there any typicality argument that suggests 
that there will be a Cauchy hypersurface that represents 
a turning point for the {\em whole} universe?
We do not see any such argument. Of course, in a huge 
eternal universe, it is not unreasonable to expect 
small spacetime regions where the entropy attains
a {\em local} minimum. However, we do not see a reason 
why such a region should take a form of a complete 
Cauchy hypersurface for the whole universe.

Apparently, Carrol and Chen have introduced the notion
of a Cauchy hypersurface with a minimal entropy 
in order to have a manifest symmetry between the two 
opposite directions of time. Does it mean that our 
criticism suggests that there is no such symmetry?
Fortunately, the answer is no! According to our 
picture, typical points in spacetime have a time arrow
in one direction only, but we cannot predict 
which direction is that, because all directions are, 
{\it a priori}, equally probable. However, the point is 
that the universe simply {\em must} choose some direction
at (almost) every point, because otherwise it will be 
a rather atypical universe. In other words, 
the choice of the direction of a local time arrow 
is a matter of {\em spontaneous symmetry breaking}.
Different regions of the universe may have different directions
of the time arrow, in an analogous (although not equivalent) 
way as they may have different 
directions of the Higgs field. 

In our observable part of the universe, the existence of a time 
arrow seems to be a rather global (not local) phenomenon.
On the other hand,
our typicality argument above by itself does not explain 
why the time arrow seems to be so uniform on such a large scale.
However, here it is the spontaneous inflation discussed 
in \cite{car} that may solve the problem. 
Owing to the inflation, an initial local 
time arrow will naturally become a large scale phenomenon. 

At the end, let us emphasize once again that, in \cite{car}, 
the existence of an infinite number of the degrees of 
freedom in an arbitrarily small patch of space is the 
crucial assumption needed for the explanation of the 
time arrow. While this assumption is certainly problematic 
from the point of view of ultraviolet divergences in 
field theory, it is interesting to note that this 
assumption may also be helpful in solving the 
black-hole information paradox \cite{nik2}. 

To conclude, we believe that the idea based on an infinite 
number of the degrees of freedom and spontaneous inflation 
introduced in \cite{car} is very promising in explaining 
the thermodynamic time arrow. However, we also believe
that our comment represents an important refinement of 
this idea.

\section*{Acknowledgment}

This work was supported by the Ministry of Science and Technology of the
Republic of Croatia under the contract No.~0098002.

\end{document}